\documentclass[twoside]{article}

\input oejv.sty

\begin{document}

\OEJVhead{March 2012}
\OEJVtitle{xNew variable stars in the field of open cluster NGC188}
\OEJVauth{Popov, A.A.$^1$; Krushinsky, V.V.$^1$; Avvakumova, E.A.$^1$; Burdanov, A.Y.$^1$; Punanova, A.F.$^1$; Zalozhnih, I.S.$^1$}
\OEJVinst{Astronomical Observatory of Ural Federal University, \\
ul. Mira d. 19, g. Yekaterinburg, Russian Federation, 620002, \\
{\tt apopov66@gmail.com}}

\OEJVabstract{A photometric study of variable stars in the field of old open cluster NGC~188 is discussed. Observations were carried out in two bands $R$ and $I$ for 5513 stars up to $R~=~17^m$ in the field of $1.5~\times~1.5^{\circ}$ around the cluster. The photometric data were processed by the console application ''Astrokit'', which corrects brightness variations associated with the variability of atmospheric transparency and carries out searching for  variable stars. We found 18 new variable stars and determined the parameters of one previously known variable. Among discovered stars one is a low-amplitude pulsating variable, one is a EW eclipsing binary, six are eclipsing variables of EA type, five objects are long period variables, and for five stars variability type remains uncertain.}

\begintext



\section{Introduction}

Study of variable stars makes a major contribution to our knowledge
of structure and evolution of stars and stellar systems. Variable stars provide a unique opportunity to determine many important characteristics, either of the stars themselves (size, mass, luminosity) or of the structure of the Galaxy (stellar population analysis). Study of variable stars in clusters is of particular importance because of two main reasons: (1) the information about distance and age of cluster can be improved using some types of variable stars like Cepheids; (2) we can determine age, distance and evolutionary stage of variable stars, which belong to clusters with known distance and age.

NGC~188 (RA2000 = 00$^h$47$^m$28$^s$, Dec2000 = +85$^{\circ}$15$'$18$''$) is well-studied cluster, what is confirmed by the large number of articles (over 500 in the last 50 years). Fornal et~al. (2007) give an adequate review of studies of the cluster fundamental parameters in last 45 years. The median values of these parameters are taken from tab.~1 of Fornal et~al. (2007): true distance modulus $11^m.25$, corresponding to the distance $1660$ pc, Age$ = 7$ Gyr, reddening E(B-V)$ = 0^m.09$ and metallicity [Fe/H]$ = 0$. The cluster was studied in a wide range of wavelengths from infrared to X-ray (see e.g. Bonatto et~al. (2005), Eggen and Sandage (1969), Janes~(1979), Sarajedini~et~al.~(1999) , McClure~(1974), Twarog~et~al.~(1997), Gondoin~(2005)). Platais~et~al.~(2003) compiled a catalogue of positions and proper motions of 7812 stars brighter $V$ = $21^m$ in the field of 0.72 deg$^2$ around the cluster center. There are several papers devoted to the search and study of variable stars in the central part of the cluster (Kaluzny~anf~Shara~(1987), Zhang~et~al.~(2002, 2004), Kafka~and~Honeycutt~(2003), Mochejska~et~al.~(2008)). In 2004 Stetson~et~al. published the photometric catalogue for a few thousand of cluster stars. We used this data (Stetson~et~al.~(2004)) as a photometric standard for transformation of instrumental system of our telescope to standard bands. Previously unknown variable stars have been found during the execution of this work. They are discussed in this paper.

\section{Observations and Data Reduction}
\begin{figure}[h]
\begin{center}
\epsfig{file=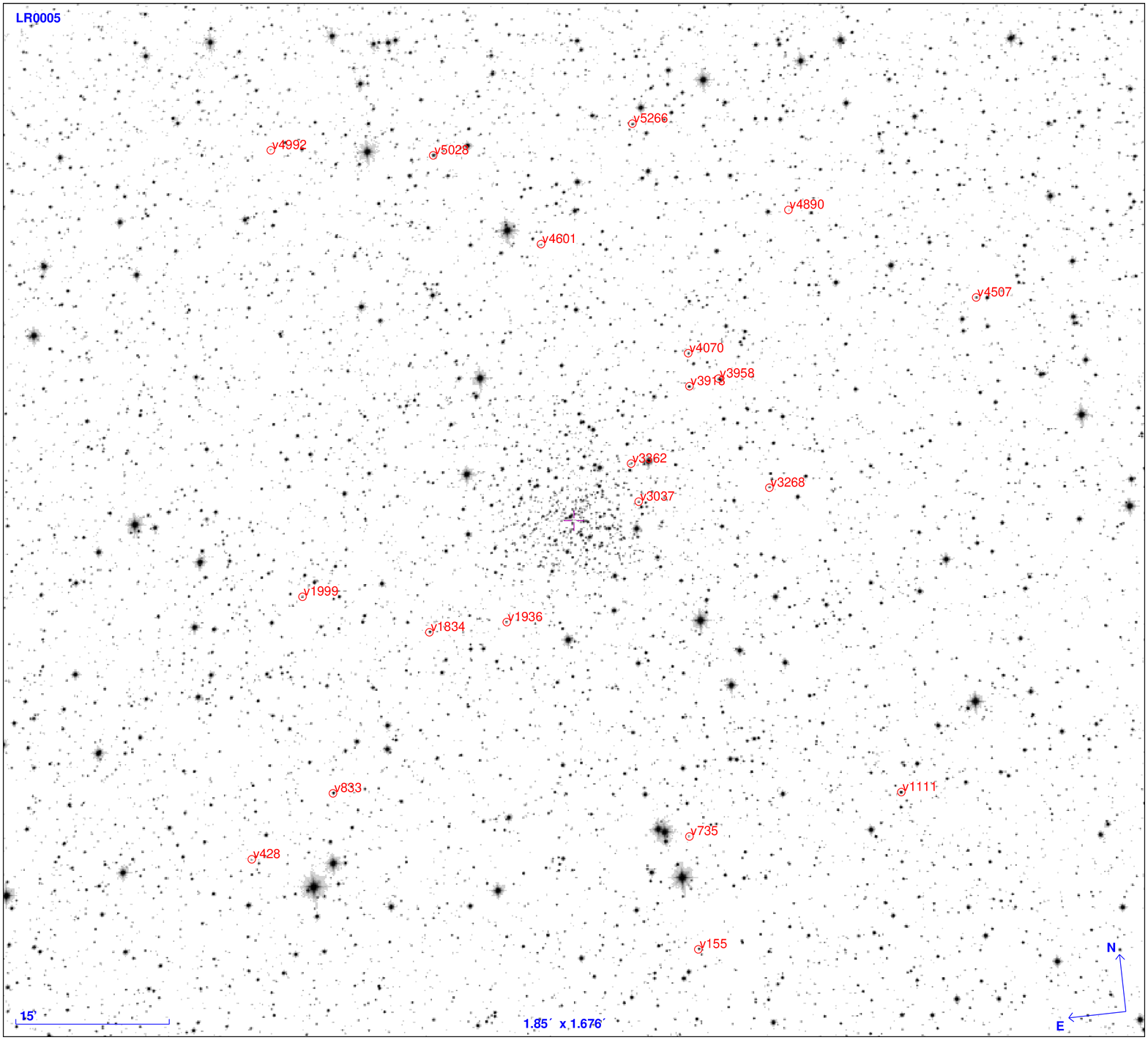, width=15 cm, keepaspectratio}
\end{center}
\caption{Observed field of open cluster NGC~188. The abscissa is a right ascension, the ordinate is a declination. Red circles denote variable stars studied in this work. This image was made using Aladin software (Bonnarel~et~al.~(2000))}
\label{ris:pole}
\end{figure}

Photometric observations of open cluster NGC~188 were performed in March 2011 in the Astronomical observatory of Ural Federal University. We used MASTER series robotic telescope (Lipunov~et~al.~(2010)). The telescope has two parallel mounted identical tubes of the Hamiltonian system with a diameter of 40~cm and the focal length of 100~cm. Apogee Alta U16M 4096x4096 Peltier-cooled CCD with a pixel size of 9x9~$\mu m$ is installed  in the main focus of each tube. Image scale is 1.85 arcsec pixel$^{-1}$ and field of view is $2\times2^{\circ}$ (see fig.~\ref{ris:pole}). The telescope is equipped with Johnson-Cousins $BVRI$ filters. Two tubes allow to observe the same object in two filters simultaneously. The observations were carried out during the period from 11 to 19 March 2011 during 5 nights. Full list of observational nights, used filters and exposure time (TE) are given in the tab.~\ref{tab:Dateofobserv}. The average seeing was 5$''$. Flat field frames were obtained from the twilight sky.

\begin{table}[h]
\begin{center}
\caption{Journal of observations}
  \begin{tabular}{c c c c}\hline
Date&$R$&$I$&TE, sec.\\\hline
11.03.11&53&27&180\\
14.03.11&37&65&180\\
17.03.11&45&45&180\\
18.03.11&21&62&180\\
19.03.11&51&15&180\\\hline
Total&207&214&\\\hline
 \end{tabular}
\label{tab:Dateofobserv}
\end{center}
\end{table}

\begin{figure}[h]
\begin{center}
\epsfig{file=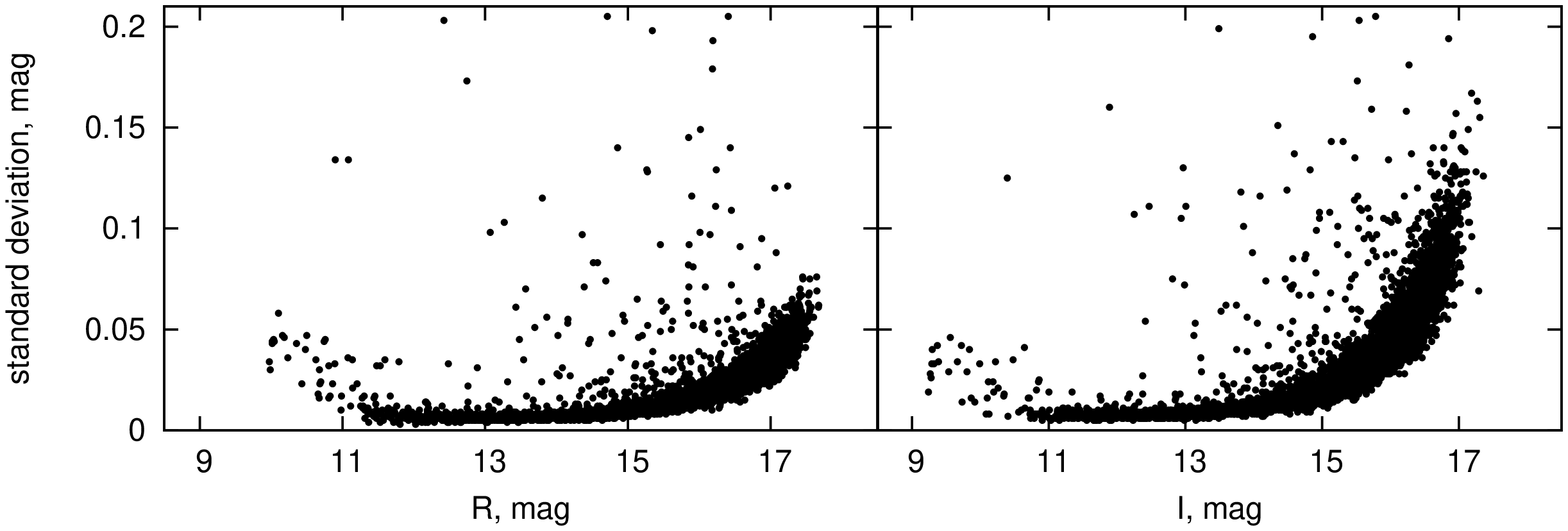, angle=-90, width=15 cm, keepaspectratio}
\end{center}
\caption{The dependence of the standard deviation of the magnitude for the bands $R$ (left) and $I$ (right)}
\label{ris:stdv}
\end{figure}

The obtained data were processed with IRAF V2.14. software package (Tody~(1986)). Processing includes the initial photometric reduction, astrometric reduction and aperture photometry. The initial reduction consists of removing overscan, subtraction of dark frames and division by normalized flat frames. Astrometric reduction was made with IRAF/images/imcoords. Tycho-2 catalogue (Hog et~al.~(2000)) was used as a reference catalogue. To carry out aperture photometry we used list of equatorial coordinates for stars obtained from the best frame. So the same star in all frames in all filters had the same ID and further processing of the photometric data became easier. The aperture photometry was carried out simultaneously for all stars from the list on all frames with the same aperture. The diameter of aperture was equal to 3 pixels ($6''$). Sky background was measured in a ring with width of 5 pixels (10$''$) and inner radius of 7 pixels ($14''$). Photometry for 5513 stars in the range of magnitudes from 11 to 17 with a corresponding accuracy from $0.006$ to $0.05$ mag (in band R) was carried out in field of $90\times90'$. Fig.~\ref{ris:stdv} shows the standard deviation of the magnitude versus the magnitude for the bands R (left) and I (right).

\begin{figure}[h]
\begin{center}
\epsfig{file=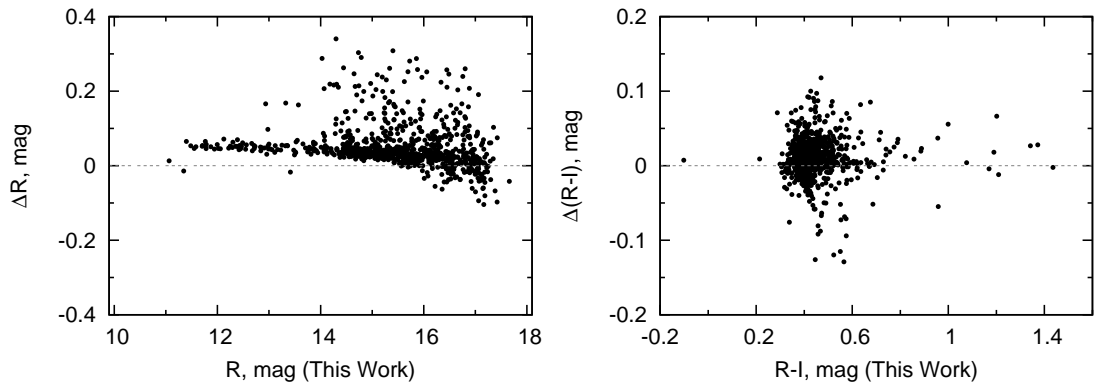, angle=-90, width=15 cm, keepaspectratio}
\end{center}
\caption{$\Delta R = R_{Sarajedini} - R_{our}$, $\Delta (R-I) = (R-I)_{Sarajedini} - (R-I)_{our}$}
\label{ris:dR}
\end{figure}

To get and to correct the average magnitudes we used a console application ''Astrokit''\footnote{The code is available by request to burdanov.art{@}gmail.com} written by two of us (Krushinsky V. and Burdanov A.) on the basis of modified algorithm of Everett~and~Howell~(2001). This application allows to correct brightness variations associated with the variability of atmospheric transparency and carries out searching for variable stars with the RoMS algorithm of Rose~and~Hintz (2007). ''Astrokit'' allows to process a large number of objects simultaneously in an automatic mode. To convert the instrumental magnitudes to the standard ones we took photometric data obtained previously for this area by Stetson~et~al.~(2004). We used linear equations~\ref{eq1} with coefficients estimated by least squares method for 790 common stars.

\begin{equation}
\begin{array}{lr}
 R-I = 0.832(r-i) + 0.719,&\sigma = 0.022,\\
 R-r = -0.163(R-I) - 0.252,&\sigma = 0.036,
\end{array}\label{eq1}
\end{equation}

where R and I are standard magnitudes, r and i are instrumental magnitudes, $\sigma$ is standard deviation of conversion.
Our photometric data are in good agreement with the results obtained by Sarajedini~et~al.~(1999). The difference between our data and Sarajedini is $\Delta (R-I) = 0^m.009\pm0^m.027 $, $\Delta R = 0^m.038\pm0^m.058$, individual residuals are shown in fig.~\ref{ris:dR}.

\begin{figure}[h]
\begin{center}
\epsfig{file=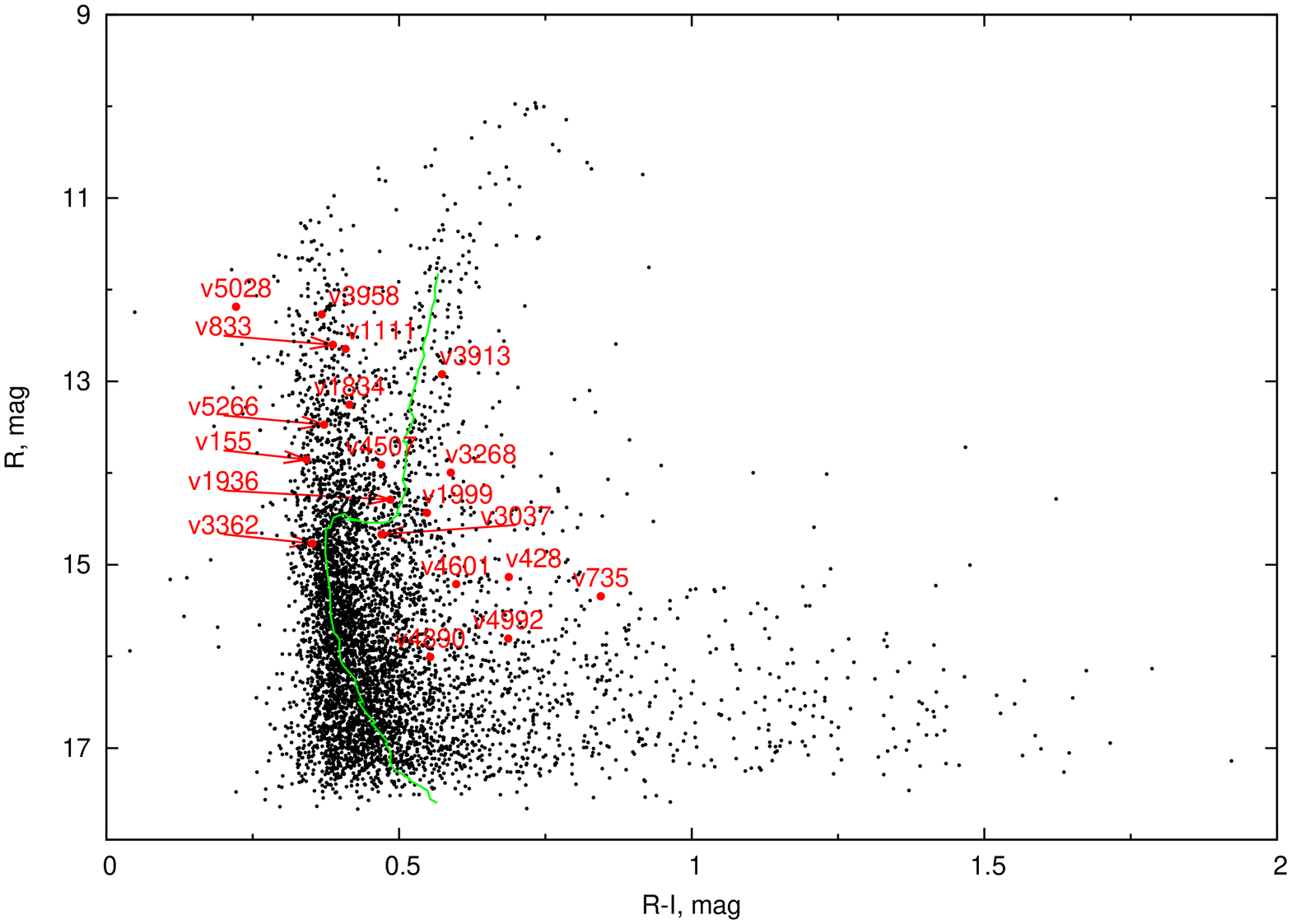, angle=-90, width=15 cm, keepaspectratio}
\end{center}
\caption{Color-magnitude diagram of NGC~188 based on our photometric data. The red circles denote new variable stars. The solid green line corresponds to an average cluster sequence by Stetson~et~al.~(2004)}
\label{ris:GR}
\end{figure}

\section{Variable stars}

Kafka~and~Honeycutt~(2003) reported about 51 variable stars in the field of NGC~188, only three of them were previously known. However Mochejska~et~al.~(2008) identified 32 stars out of these 48 variables and only for one they found brightness variations.

Today 58 variable stars are known within 2$^{\circ}$ radius around NGC~188 according to the electronic catalogue of the American Association of Variable Star Observers (AAVSO/VSX). We didn't investigate the known variable stars if our data neither contradict to the previous one nor clarify them. So we refined the parameters for one previously known variable and discovered 18 new ones. Tab.~\ref{tab:varstars} shows the parameters of the studied variable stars. The table contains the identification number (ID) of the star in our list, 2MASS identification number (Skrutskie~et~al.~(2006)), the equatorial coordinates, the observed maximum and minimum magnitudes in the filters R and I (whenever it is possible), the variability type, the ephemeris (if it can be determined from our data) and the probability of membership (MP) in the cluster. The estimates of the membership probability are taken from Platais~et~al.~(2003). Positions in cluster color-magnitude diagram (CMD) of all discovered variables relative to the average NGC~188 sequence of Stetson~et~al.~(2004) are shown in  fig.~\ref{ris:GR}. We use this positions for some speculations about membership for our variables hereinafter, because we didn't determine properly which of our stars are really members of NGC~188. Times of minima  were obtained with Kwee and van Woerden's (1956) method in ''AVE''\footnote{http://www.astrogea.org/soft/ave/aveint.htm} software when our data allow such determination. Derived values for one or both filters are listed in  tab.~\ref{tab:minima}. To estimate the periods of variable stars we used the ''Winefk''\footnote{http://vgoray.front.ru/software/} software which realise the method of periodogram analysis. The software was developed by Goransky~V.~P. We managed to found  periods only for seven variables from our list.

\begin{table}[h]
    \setlength{\tabcolsep}{2pt}
\begin{small}
\begin{center}
\caption{Parameters of new variable stars}
  \begin{tabular}{c c c c c c c c c c c c c}\hline
ID   &2MASS           &$\alpha_{2000}$&$\delta_{2000}$    &\multicolumn{3}{c}{R, mag}&\multicolumn{3}{c}{I, mag}&Type       &Ephemeris&MP\\
     &                &$h$ $m$ $s$    &$^{\circ}$ $'$ $''$&max   &minI  &minII       &max   &minI  &minII       &           &JD-2455000&\%\\\hline
v155 &00365525+8433366&00 36 55       & +84 33 36         &13.853&14.018&13.99       &13.436&13.604&13.573      &EW         &640.2628+0.3518E&  \\
v428 &01081272+8438061&01 08 12       & +84 38 06         &14.971&15.675&15.552      &14.382&15.039&14.952      &EB         &632.2605+0.3077E&  \\
v735 &00375107+8444311&00 37 51       & +84 44 31         &15.294&15.478&            &14.455&14.554&            &VAR        &       &  \\
v833 &01031463+8445435&01 03 14       & +84 45 43         &12.669&12.695&            &12.196&12.219&            &EA:        &       &  \\
v1111&00224443+8448486&00 22 44       & +84 48 48         &12.706&12.914&            &12.130&12.351&            &VAR        &       &  \\
v1834&00574795+8502290&00 57 47       & +85 02 29         &13.267&13.346&            &12.794&12.867&            &VAR        &638.2537+0.8416E& 0\\
v1936&00520870+8504146&00 52 08       & +85 04 14         &14.357&      &            &14.894&      &            &EA         &       & 0\\
v1999&01073954+8504072&01 07 39       & +85 04 07         &14.450&14.616&            &13.787&13.922&            &L:         &       &  \\
v3037&00424049+8516494&00 42 40       & +85 16 49         &14.731&14.851&            &14.151&14.256&            &LB         &       &98\\
v3268&00322120+8518390&00 32 21       & +85 18 38         &13.914&14.083&            &13.255&13.409&            &L:         &       &34\\
v3362&00432396+8520325&00 43 23       & +85 20 32         &14.784&14.863&            &14.365&14.439&            &VAR        &632.3285+0.3102E&93\\
v3913&00390088+8528154&00 39 00       & +85 28 15         &12.989&13.052&            &12.336&12.381&            &L:         &       &39\\
v3958&00363416+8529046&00 36 34       & +85 29 04         &12.336&12.360&            &11.891&11.920&            &EA         &       & 0\\
v4507&00145705+8536311&00 14 57       & +85 36 31         &13.870&13.991&            &13.359&13.471&            &LB         &       &  \\
v4601&00521410+8541065&00 52 14       & +85 41 06         &15.226&15.342&            &14.558&14.662&            &VAR        &640.5000+2.2555E& 0\\
v4890&00305224+8545368&00 30 52       & +85 45 36         &15.978&16.363&            &15.433&      &            &EA         &       &  \\
v4992&01164582+8546202&01 16 45       & +85 46 20         &15.889&16.601&            &15.111&15.884&            &EA         &635.3397+0.4017E&  \\
v5028&01022872+8548314&01 02 28       & +85 48 31         &12.245&12.281&            &11.947&11.975&            &$\delta$Sct/$\beta$Cep&635.2187+0.0665E&  \\
v5266&00450068+8553285&00 45 00       & +85 53 28         &13.545&      &            &13.093&13.268&            &EA         &       &  \\\hline
 \end{tabular}
\label{tab:varstars}
\end{center}
  \end{small}
\end{table}

\begin{table}[h]
    \setlength{\tabcolsep}{2pt}
\begin{small}
\begin{center}
\caption{Times of minima for nine discovered variables}
  \begin{tabular}{c c c c c}\hline
ID&JD(2455...)&Error&Type&Filter\\\hline
v155&632.3428&0.0006&sec&I\\
&632.3432&0.0005&sec&R\\
&635.3362&0.0008&pri&I\\
&640.2588&0.0015&pri&R\\
v428&632.2584&0.0005&pri&R\\
&635.3352&0.0002&pri&I\\
&638.2593&0.0011&sec&R\\
&638.2565&0.0009&sec&I\\
&639.3325&0.0004&pri&I\\
&640.2480&0.0026&pri&R\\
v833&635.2535&0.0014&-&R\\
v1999&635.2622&0.0004&-&R\\
&635.2891&0.0020&-&I\\
v3268&635.2581&0.0004&-&I\\
&365.2546&0.0014&-&R\\
v3362&632.3324&0.0006&-&I\\
&632.3301&0.0012&-&R\\
&638.2674&0.0057&-&I\\
&638.2534&0.0004&-&R\\
v4992&635.3384&0.0005&pri&I\\
&639.3538&0.0087&pri&I\\
v5028&632.2926&0.0009&-&R\\
&635.2846&0.0017&-&R\\
&638.3400&0.0031&-&R\\
&640.2695&0.0011&-&R\\
&640.3282&0.0006&-&R\\
v5266&635.3554&0.0054&-&I\\\hline
 \end{tabular}
\label{tab:minima}
\end{center}
  \end{small}
\end{table}

For classification of some discovered variable stars we used the additional photometric data from the 2MASS catalogue, since the infrared interstellar extinction is minimal. Pickles~(1998) gives the values of the spectral type and corresponding color indices. We estimated approximate spectral classes for discovered variables using three color indices (R-I)c, (J-H) and (H-K),which are listed in column~5 of tab.~\ref{tab:colors}. We calculated upper limits of the distances for stars based on their proper motions from the catalogue PPMX (Roeser~et~al.~(2008)) using the equation

\begin{equation}
\begin{array}{lr}
Dist\mu=Vt/4.74\surd(\mu_\alpha^2+\mu_\delta^2),
\end{array}\label{eq2}
\end{equation}

where $Vt$ is 200~km/s (mean speed of stars in the Galaxy disk). Derived values are given in column~6 of tab.~\ref{tab:colors}. In accordance with the size of the Galaxy the distance does not exceed 15~kpc.

Lower limit of the distances was obtained under the assumption that all our variables are the main sequence stars. We calculated the distances using the absolute magnitudes in the band J (Mj) for the main sequence stars of different spectral types given by Pickles~(1998).

\begin{equation}
\begin{array}{lr}
Dist\_Mj\_V=d_010^{((J-Mj)/5)},
\end{array}\label{eq3}
\end{equation}

where $d_0$ is 10~pc. These values ​​are given in column~7 of table~\ref{tab:colors}.

\begin{table}[h]
    \setlength{\tabcolsep}{2pt}
\begin{small}
\begin{center}
\caption{Colors of new variable stars}
  \begin{tabular}{c c c c c c c c}\hline
ID   &R-I  &J-H  &H-K  &Sp&Dist$\mu$&Dist\_Mj\_V&Class     \\\hline
v155 &0.342&0.292&0.007&G1&6786     &968        &V-III     \\
v428 &0.687&0.461&0.124&K1&11193    &784        &V-III     \\
v735 &0.845&0.683&0.145&K7&815      &349        &V     \\
v833 &0.387&0.34 &0.084&G8&1889     &365        &V-IV     \\
v1111&0.409&0.42 &0.097&K0&18276    &288        &V-II     \\
v1834&0.415&0.395&0.053&G8&2509     &496        &V-III     \\
v1936&0.485&0.487&0.074&K2&3771     &532        &V-III     \\
v1999&0.548&0.572&0.171&K4&8101     &360        &V-III     \\
v3037&0.47 &0.539&0.116&K2&11100    &631        &V-III     \\
v3268&0.588&0.562&0.216&K4&3127     &284        &V-IV     \\
v3362&0.352&0.359&0.041&G2&4417     &1283       &V-IV     \\
v3913&0.573&0.575&0.15 &K4&9750     &189        &V-III     \\
v3958&0.368&0.295&0.079&G1&7487     &433        &V-III     \\
v4507&0.469&0.385&0.055&K0&2795     &525        &V-III     \\
v4601&0.598&0.502&0.186&K4&2565     &503        &V     \\
v4890&0.553&0.377&0.267&K4&47174    &839        &V-III     \\
v4992&0.687&0.66 &0.138&K7&2391     &460        &V     \\
v5028&0.221&0.163&0.047&F2&5020     &657        &V-III     \\
v5266&0.372&0.271&0.058&G0&10935    &758        &V-III     \\\hline
 \end{tabular}
\label{tab:colors}
\end{center}
  \end{small}
\end{table}

Comparing the distances computed by two methods we can say that all the studied stars are not closer than 200~pc. Futher we suggested that all our stars are subgiants and estimated the new values of distances to the stars Dist\_Mj\_IV using the eq.~\ref{eq3} again. As a result the distances Dist\_Mj\_IV to the stars ID v735, v4601 and v4992 appear to be larger than Dist$_\mu$. Therefore we can suggest that variables with ID v735, v4601 and v4992 are the main sequence stars indeed. Additionally in the same way we estimated the distances to the variables in cases they are giants and supergiants. If new value of distance derived for some star is larger than $Dist\mu$ then this star couldn't be giant or supergiant respectively. The final values of luminosity classes are listed in column~8 of tab.~\ref{tab:colors}.

Such estimation allows us to specify not only spectral types for our variables but roughly determine the location of each variable on the CMD. Moreover estimates of spectral type and distance from photometric data are useful when variability types are defined.

Observations and results for each star are discussed in the next section.

\section{Discussion}

\subsection*{v155}
The light curve of v155 is presented on the left top panel of fig.~\ref{ris:period}. We derived four times of minima for this variable and were able to determine the period. Because of the shape of the light curve we decided that it is eclipsing binary system of EW type. The depth of the primary minimum equal to about 0$^m$.17 for R and I bands. The depth of the secondary minimum is about 0$^m$.14 in both filters. As can be seen from our light curve the secondary minimum is the total eclipse with the duration of about 0.054P and 0.096P in R and I filters respectively. The precision of our observations is not suitable for the better determination. Nevertheless the almost equal depth of minima, continuous light variations outside the eclipses, short period and spectral type point out to a rather contact or near contact system.

Unfortunately  Platais~et~al.~(2003) didn't investigate the area with v155 so we can't say anything about membership in NGC~188 for this system except that its position relative to the average NGC~188 sequence of Stetson~et~al.~(2004) (see fig.~~\ref{ris:GR}) points out that it is likely field star.

\subsection*{v428}
Star v428 is denoted in the VSX database as suspected variable NSV395 based on the old photographic data (Hoffmeister, 1964). We checked also Simbad database and didn't find any kind of information about this star. So our observations are the first CCD light curve for NSV395 and it is presented on the right top panel of fig.~\ref{ris:period}. It covers six minima in both filters what is enough for satisfactory determination of period and type of variability.

As one can see the shape of the light curve resembles ones for the eclipsing binary systems denoted usually as EB. As in the case of v155 the secondary eclipse is total with duration is about 0.08P in R filter. The totality of the secondary eclipse in I filter is less obvious. The depth of the primary minimum equals to about 0$^m$.70 in R band and to 0$^m$.66 in I band.  The depth of the secondary minimum is about 0$^m$.58 and 0$^m$.57 in R and I filters respectively. The difference between  values given for R and I bands can be explained by the presence of the small light fluctuations with amplitude of about 0$^m$.06 during the both minima on the light curve at least in the R filter.

Based on the position of this star in relation to cluster CMD we assume that v428 is a field star.

\subsection*{v735}
The light curve of v735 is shown on the left panel of fig.~\ref{ris:unknown}. During our observations we only registered slow increase of brightness from 17 till 19 of March  with amplitudes of about 0$^m$.10 and 0$^m$.18 in the I and R filter respectively. For the first two nights namely March, 11 and 14 brightness of star also increased. We didn't collect any times of minimum or maximum and couldn't determine both period and the character of light variations.

There is no value of membership probability for v735 in the work of Platais~et~al.~(2003) and Geller~et~al.~(2009), but star is far away from average cluster sequence of Stetson~et~al.~(2004) and cluster center so we can assume that it is not the member of NGC~188.

\subsection*{v833} 
The light curve of this variable is presented on the left top panel of fig.~\ref{ris:EA}. During our observational set only one minimum was detected in R band on March, 14. We tried to determine also time of minimum in I filter for the same night, but observational points were collected some later than minimum occures. However light variations are obvious from our data: the brightness was almost constant (within the observational accuracy) on March, 11 and March 17-19, while on March, 14 brightness decreased in R band notably. In a less degree this effect is seen on the light curve in I filter. We suppose that such brightness variations indicate an eclipsing binary with EA light curve, but more continuous set of observations is required to confirm or reject our suggestion.

As in the case of v155 and v428 this variable lies far away from cluster CMD and outside the region investigated by Platais~et~al.~(2003) thus we suppose it is rather a field star.

\subsection*{v1111}
\begin{figure}[h!]
\begin{center}
\epsfig{file=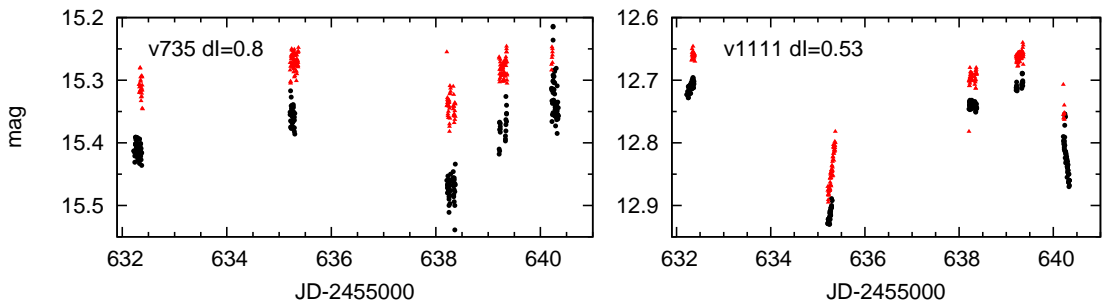, angle=-90, width=15 cm, keepaspectratio}
\end{center}
\caption{Variable stars with no measurable light curves classification. Black circles are band R, red triangles are band I. dI indicates a shift of magnitude in band I}
\label{ris:unknown}
\end{figure}

The light curve of v1111 is shown on the right panel of fig.~\ref{ris:unknown}. We registered slow and small increase of brightness during March, 11 and 14 in both R and I filters and small decrease of brightness during March, 19 in the R band only. In the I filter observations on 19 of March were started some later and brightness was almost constant. Neither moments of minima nor maxima had been observed thus we couldn't determine the period and classify this variable.

There is no information about value of membership probability because star was not included to the work of Platais~et~al.~(2003) and Geller~et~al.~(2008). Due to its position relative to the cluster center and CMD we assume that it is a field star.

\subsection*{v1834}
Star v1834 shows sinusoidal light variations, its light curve is presented on the left middle panel of fig.~\ref{ris:period}. We observed slow decrease of brightness on March, 14 from about 12$^m$.81 to 12$^m$.88 and slow increase of brightness from about 12$^m$.86 to 12$^m$.80 on March, 17 till March, 19  in I band. The light variations in R filter were similar. We suppose that at least one minimum had to occur between 14 and 17, but we didn't carry out the observations because of the bad weather. As we didn't cover times of minima, the value of period determined with ''Winefk'' software and given in tab.~\ref{tab:varstars} for this star should be considered only as initial and rough guess.

Spectral type of v1834 derived in this work is G8V--III. Platais~et~al.~(2003) had given zero membership probability for this star, Geller~et~al.~(2008) in their investigation of the  stellar radial velocities NGC~188 also couldn't find the MP value for v1834. That is why we could not determine more confidently the position on the CMD based on Stetson~et~al.~(2004) data.   So we only believe that it is a star of middle spectral type. In this case its variability can be caused by the presence of fast rotation or/and cool spots. Geller~et~al.~(2008) classified this star as a rapid rotator based on their RV measurements. This fact confirmes our hypothesis.

\subsection*{v1936}
The light curve of star v1936 is shown on the right top panel of fig.~\ref{ris:EA}. It is obvious that during four nights the brightness of this object was almost constant with small fluctuations around the values listed in tab.~\ref{tab:varstars}. However on March, 19 we observed slow increase of the brightness during all night. The light changes in R band were from 14$^m$.44 to the maximum level (14$^m$.36). Due to fact that the observation in the I filter started later and lasted shorter we registered smaller increase of brightness from 15$^m$.04 to 14$^m$.94 only. Decrease of the brightness had to occur between two observational nights, but we didn't observe moment of minimum and thus we can suppose only that depth of the minimum is greater than 0$^m$.1. Such fast decreasing and increasing of the light followed by long-time constant brightness can point out to the eclipsing binary of EA type.

As can be seen from fig.~\ref{ris:GR} v1936 occupies relevant position in cluster CMD close to the subgiant branch, although the MP value given by Platais  equals to 0 and Geller~et~al.~(2008) didn't include star in their work.

\subsection*{v1999}
Star v1999 shows slow changes in brightness resembling long-period giant variables. Its light curve is presented on the left top panel of fig.~\ref{ris:L}. We observed only one minimum on March, 14. During three last nights after the brighntess of the star slowly increased. We can say nothing about the amplitude of light variations because maximum of brightness wasn't registered.

v1999 lies far away from cluster center and was not included in the study of Platias thus nothing is known about its membership in the NGC~188. However it is close to the red giant branch of cluster CMD. Our preliminary spectral type (see tab.~\ref{tab:colors}) is K4V--III so v1999 can be the late type giant star indeed. Thereby we decide to denote its type variability as 'L:'.

\subsection*{v3037}
The light variations of v3037 are similar to ones for previous star v1999. The derived light curve is shown on the top right panel of fig.~\ref{ris:L}. We had not observe moment of minimum, but it is clearly seen from light curves that it occurred between March, 14 and March, 17. Furthermore the moment of maximum brightness is undetermined too. The registered decrease of light between March,11 and March,14 is about 0$^m$.1 in both bands. Then during three nights brightness of the star continuously increased by 0$^m$.12. So one can suspect that amplitude of variations is not less than these values.

This star is a X-ray source according to Gondoin (2005), who also determined spectral type as G8III based on (B-V) color index.

The membership probability is known for this star. We listed MP value according to Platias work, while Stetson gives smaller value namely 78\% based on average data from Platias and Dinescu~et~al.~(1996). The position of v3037 relative to the cluster sequence (see fig.~\ref{ris:GR}) and our estimate of spectral type (K2V--III) both can be used as evidence that this star is a late--type subgiant or giant with slow light variation caused by pulsations. But X-ray emission can points out to the rapid rotation and magnetic activity of this star (Gondoin, 2005). As we did not determine the period of photometric variations we use type 'LB' as a preliminary.

\subsection*{v3268}
The light curve of v3268 we derived is shown on the left middle panel of fig.~\ref{ris:L}. Our observations cover one moment of minimum in both filters. The amplitude of registered light changes equals to 0$.^m$16 in R band and to 0$^m$.15 in I band.  The shape of the light curve looks like light curve of the pulsating variables.

Geller~et~al.~(2008) classified v3268 (ID 3118 for their study) as double-lined spectroscopic binary and published (see Geller~et~al., 2009) radial velocity curves for both components. Geller~et~al.~(2009) also determined orbital period P=11$^d$.9022 and masses for both components. We tried to determine period from our data but unsuccessfully. One could expect the variability caused by eclipses, but orbital period found by Geller~et~al.~(2009) disagrees with photometric brightness variations.

The membership probability for v3268 was determined by several authors. While Platais~et~al.~(2004) gave small value for MP (see tab.~\ref{tab:varstars}), Stetson's estimate based on values from Platais~et~al.~(2003) and Dinescu~et~al.~(1996) was higher and equaled to 48\%, recently Geller~et~al.~(2008) estimated MP to be 88\%.  We estimate spectral type of v3268 as K4V-IV, but system lies far from main sequence of NGC~188 so it is rather an evolved star and its photometric variability is caused by pulsations.

\subsection*{v3362}
The light curve of v3362 is presented on the middle right panel of fig.~\ref{ris:period}. Our observations covered two minima in each bands, so we were able to determine the period value. With derived period v3362 shows sinusoidal light variations with amplitude of about 0$^m$.11 in both filters.

The value of the membership probability was given for this star by Platais and Stetson, both values are sufficient for v3362 to be a member of NGC~188.  Geller~et~al.~(2008) had not determine MP value from their radial velocity study, but classified v3362 as one-lined spectroscopic binary and rapid rotator, but  radial velocity curve was not published in their next paper (Geller~et~al., 2009). Gondoin (2005) included this star to his list of X-ray sources, but obstructed to determine the reason of X-ray emission.

We assumed the G2V--IV spectral type, but because the position of the v3362 is close to the main sequence of NGC~188 it is rather G2V, thus its variability can be caused by rotation or/and chromospherical activity and cool spots.

\subsection*{v3913}
The light curve of the v3913 is shown on the middle right panel of fig.~\ref{ris:L}. As can be seen, we observed slow decreasing of the brightness during all 5 nights with amplitude of about 0$^m$.1 in both I and R bands. Without observed moments of minima we did not determine the period of light variations.

We checked the literature and found that membership probability of Stetson is some higher than Platais gave, while neither Dinescu~et~al.~(1996) nor Geller~et~al.~(2008) estimated MP value. However last authors classified v3913 as one-lined spectroscopic binary. We should suppose only that v3913 is a member of NGC~188. In this case from fig.~\ref{ris:GR} it is clearly seen that star is close to the red giant branch of average sequence of Stetson, so considering our spectral type estimate for v3913 it is rather K4III star with slow variations of light.

\subsection*{3958}
The light curve of v3958 is presented on the left middle panel of fig.~\ref{ris:EA}. Twice (March, 11 and 17) during our observations we registered decreasing of brightness up to 0$^m$.041 in I filter and up to 0$^m$.03 in R filter. But unfortunately on March, 11 set of observations ended before minimum arises, and on March, 17, we were able to register decrease and followed by increase of brightness, but not moment of minimum itself.

The value of membership probability was given by Platais and Stetson, both values are close to zero. Also Geller~et~al.~(2008) denoted v3958 as probably SB2 system, but didn't determine the MP value. So as this star lies far from average cluster sequence on fig.~\ref{ris:GR} we can suppose that it is a field star. Based on the shape of light curve and data of Geller we classified v3958 as eclipsing binary of EA type.

\subsection*{v4507}
\begin{figure}[h!]
\begin{center}
\epsfig{file=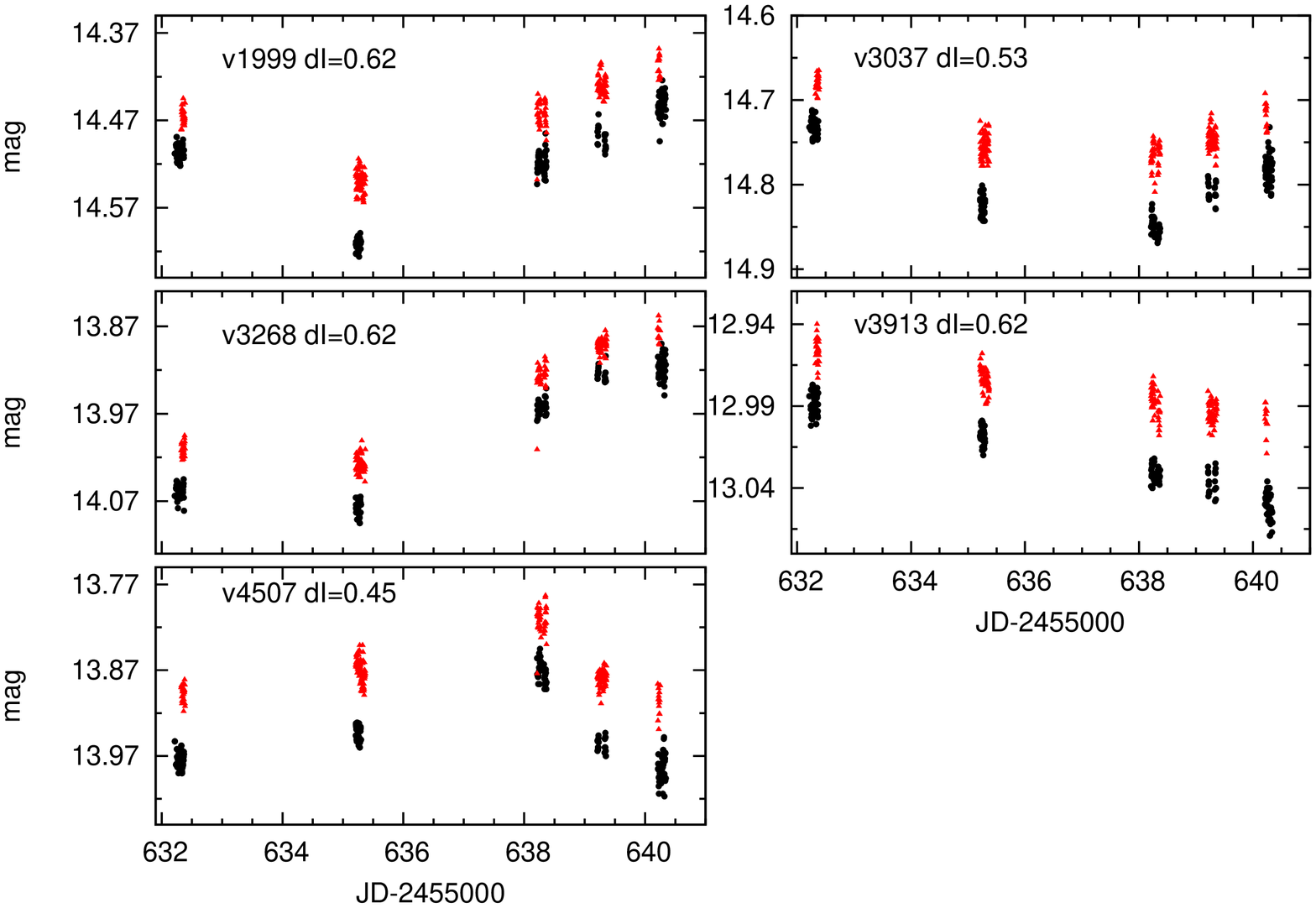, angle=-90, width=15 cm, keepaspectratio}
\end{center}
\caption{The light curves of slowly pulsating giants. Black circles are band R, red triangles are band I. dI indicates a shift of magnitude in band I}
\label{ris:L}
\end{figure}

Light curve of the v4507 is shown on the left bottom panel of fig.~\ref{ris:L}. As can be seen, we observed slow increase for three days and subsequent fast decrease of brightness.  We didn't register neither maxinum nor minimum definitely.

Also there are no estimates of membership probability for v4507, because it lies far from cluster center, however star occupies relevant position in CMD and can be the possible member of NGC~188. Due to its light curve and derived spectral type we suspect v4507 to be a giant, which variability is caused by the long period pulsations.

\subsection*{v4601}
\begin{figure}[h!]
\begin{center}
\epsfig{file=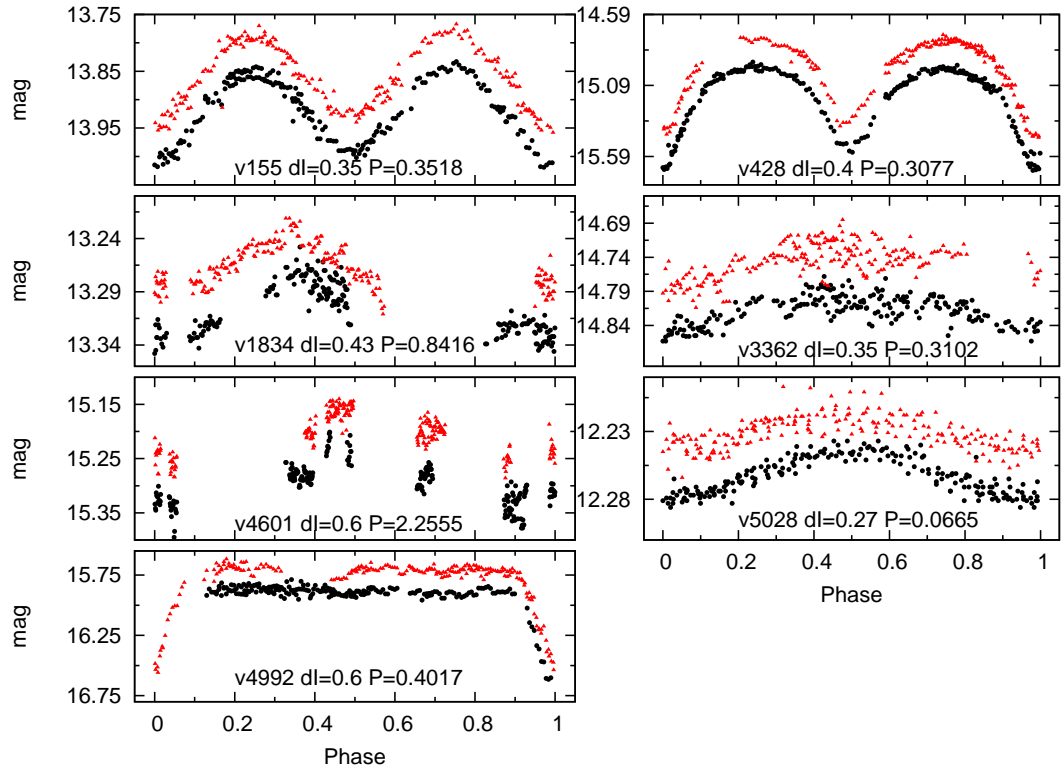, angle=-90, width=15 cm, keepaspectratio}
\end{center}
\caption{The phase curves of periodic variable stars. Black circles are band R, red triangles are band I. dI indicates a shift of magnitude in band I}
\label{ris:period}
\end{figure}

Star v4601 shows sinusoidal light variations with period listed in tab.~\ref{tab:varstars}. Its light curve is presented on the left bottom panel of fig.~\ref{ris:period}. We carried out observations around the minimum brightness twice on March, 17 and March, 19, but didn't register minimum itself. So the value of period we derived with ''Winefk'' is some preliminary.

Both Platais and Stetson gave 0\% for the membership probability of v4601. This star lies far away from average cluster sequence of Stetson et al.~(2004) (see fig.~\ref{ris:GR}) and from cluster center (see fig.~\ref{ris:pole}) therefore we suppose v4601 is not a member of the NGC~188.

According to our preliminary value of the spectral type K4V v4601 is a main sequence star thus considering its light curve we suspect that its variability can be caused by the presence of cool spots.

\subsection*{v4890}
\begin{figure}[h!]
\begin{center}
\epsfig{file=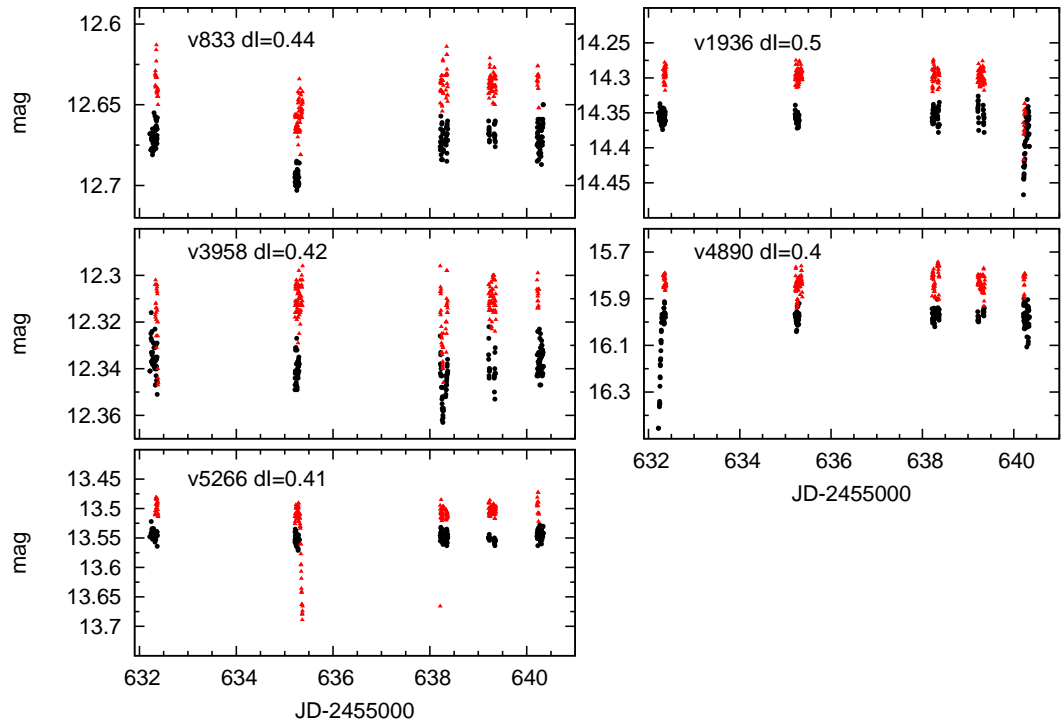, angle=-90, width=15 cm, keepaspectratio}
\end{center}
\caption{The light curves of eclipsing variable stars type of Algol. Black circles are band R, red triangles are band I. dI indicates a shift of magnitude in band I}
\label{ris:EA}
\end{figure}

The light curve of the v4890 is shown on the right middle panel of fig.~\ref{ris:EA}. During the full course of our observations the brightness of star was almost constant except on March, 11 when we registered the increase of brightness from about 16$^m$.45 to 15$^m$.98 in R filter. However our observations started some later than minimum occurs, so we couldn't determine the period.

Star v4890 is far away from the cluster center and average cluster sequence of Stetson, thus neither Platais, nor Geller~et~al.~(2008) included it to their investigations. We can suppose only, that v4890 is a rather field star. Due to the shape of its light curve we classified v4890 as an eclipsing binary of EA type.

\subsection*{v4992}
The light curve of v4992 is presented on the left bottom panel of fig.~\ref{ris:period}. We had registered moments of minima twice on March, 14 and March, 18 in I filter. The depth of the minimum was about 0$^m$.84. Observations in the R filter were shorter on both of these nights thus we didn't cover moments of minima, so our suggestion is preliminary. The brightness was almost constant during other nights. Using derived times of minima we determined the value of period. The shape of the light curve clearly points out to an eclipsing binary of EA type. It should be noted that the depth of the secondary minimum is close to zero, but small shift of moment of secondary eclipse from phase 0.5 is possible.

There is no value of membership probability for this object. Based on the position of the v4992 relative to the cluster center and average cluster sequence of Stetson we suppose that variable is rather a field star.

\subsection*{v5028}
The light curve of v5028 is shown on the right bottom panel of the fig.~\ref{ris:period}. During the course of observations we had registered several fast variations of the brightness with amplitude of about 0$^m$.04 and 0$^m$.06 in the R and I bands respectively.

Nothing is known about membership probability for v5028, because neither Platais nor Geller investigated the area in which star locates.

Such fast changes of brightness together with its short period and our estimate of spectral type (F2V--III) point to the low-amplitude pulsating variable, perhaps a $\delta$ Sct type.

\subsection*{v5266}
The light curve of variable v5266 is presented on the right bottom panel of fig.~\ref{ris:EA}. We were able to register only one moment of minimum namely on March, 14 in the I band. The observations in the R band ended before minimum arises. The depth of the minimum in the I filter equals to about 0$^m$.18. Using only one time of minimum we couldn't determine the value of period, but based on the shape of the light curve we suggest that variation of the brightness are caused by eclipses.

This star lies far away from the cluster center and from average cluster sequence thus we suppose it is rather a field star.

\section{Conclusions}
In this paper we presented results of searching for variables stars in the field of old open cluster NGC~188. CCD photometry in two bands $R$ and $I$ was made for 5513 stars up to $R = 17^m$ in the field of $1.5\times1.5^{\circ}$ around the cluster. We discovered 18 new variable stars and determined parameters of one previously known variable star (NSV395). Among discovered variable stars one is of low-amplitude pulsating variable type, one is of W~UMa type, six are eclipsing binaries of Algol type, five are long period variables, and for five stars variability type remains uncertain. Seven variables are probably cluster members. The study of open clusters in a wide field of view leads to a more complete selection of variable stars in the cluster.

\subsection*{Acknowledgements}
We thank Goransky~V.~P. (SAI MSU, Russia) for the ''WinEfk'' software and to Rafael Barbera (Grupo de Estudios Astronomicos, Spain) for his AVE software.

This work was partially supported by a grant in the form of a subsidy from Ministry of Education of the Russian Federation (the Agreement by 08.27.2012 N 8415), by the Federal task program ''Research and operations on priority directions of development of the science and technology complex of Russia for 2007-2013'' (contract 14.518.11.7064) and by Russian Foundation for Basic Research grants 12-02-31095 and 10-02-00426

This research was performed with use of Aladin (Bonnarel et al., 2000), SIMBAD database (operated at the Centre de Donn\'ees astronomiques de Strasbourg) and International Variable Star Index (VSX) database (operated at AAVSO, Massachusetts, USA).


\newpage
\begin{thebibliography}{}
\bibliographystyle{plainnat}

\bibitem[Bonatto et~al.(2005)]{Bonatto+2005}Bonatto, C., Bica, E., \& Santos, J.~F.~C., Jr., 2005, {\it AAp}, {\bf433}, 917, \OEJVbibcode{2005A\&A...433..917B}
\bibitem[Bonnarel et~al.(2000)]{Bonnarel+2000}Bonnarel, F., Fernique, P., Bienaym{\'e}, O., et~al., 2000, {\it AApS}, {\bf143}, 33, \OEJVbibcode{2000A\&AS..143...33B}
\bibitem[Dinescu~et~al.(1996)]{Dinescu+1996}Dinescu, D.~I., Girard, T.~M., van Altena, W.~F., Yang, T.-G., \& Lee, Y.-W. 1996, {\it aj}, {\bf111}, 1205, \OEJVbibcode{1996AJ....111.1205D}
\bibitem[Eggen \& Sandage(1969)]{Eggen+1969}Eggen, O.~J., \& Sandage, A.~R., 1969, {\it ApJ}, {\bf158}, 669, \OEJVbibcode{1969ApJ...158..669E}
\bibitem[Everett \& Howell(2001)]{Everett+2001}Everett, M.~E., \& Howell, S.~B. 2001, {\it PASP}, {\bf113}, 1428, \OEJVbibcode {2001PASP..113.1428E}
\bibitem[Fornal et~al.(2007)]{Fornal+2007}Fornal, B., Tucker, D.~L., Smith, J.~A., et~al., 2007, {\it AJ}, {\bf133}, 1409,\OEJVbibcode{2007AJ....133.1409F}
\bibitem[Geller et~al.(2008)]{Geller+2008}Geller, A.~M., Mathieu, R.~D., Harris, H.~C., McClure, R.~D. 2008, {\it AJ}, {\bf135}, 2264, \OEJVbibcode{2008AJ....135.2264G}
\bibitem[Geller~et~al.(2009)]{Geller+2008}Geller, A.~M., Mathieu, R.~D., Harris, H.~C., \& McClure, R.~D. 2009, {\it aj}, {\bf137}, 3743,\OEJVbibcode{2009AJ....137.3743G}
\bibitem[Gondoin(2005)]{Gondoin2005}Gondoin, P., 2005, {\it AAp}, {\bf438}, 291, \OEJVbibcode{2005A\&A...438..291G}
\bibitem[Goranskij(2000)]{Goranskij2000}Goranskij, V. P., SAI MSU, \OEJVlink{http://vgoray.front.ru/software/}
\bibitem[Hoffmeister(1964)]{Hoffmeister1964}Hoffmeister, C., 1964, {\it AN}, {\bf288}, 49, \OEJVbibcode{1964AN....288...49H}
\bibitem[Hog et~al.(2000)]{Hog+2000}Hog, E., Fabricius, C., Makarov, V.~V., et~al., 2000, {\it VizieR Online Data Catalog}, {\bf1259}, 0, \OEJVbibcode{2000yCat.1259....0H}
\bibitem[Janes(1979)]{Janes1979}Janes, K.~A., 1979, {\it ApJS}, {\bf39}, 135, \OEJVbibcode{1979ApJS...39..135J}
\bibitem[Jordi et~al.(2006)]{Jordi+2006}Jordi, K., Grebel, E.~K., \& Ammon, K., 2006, {\it AAp}, {\bf460}, 339, \OEJVbibcode{2006A\&A...460..339J}
\bibitem[Kafka \& Honeycutt(2003)]{Kafka+2003}Kafka, S., \& Honeycutt, R.~K., 2003, {\it AJ}, {\bf126}, 276, \OEJVbibcode{2003AJ....126..276K}
\bibitem[Kaluzny \& Shara(1987)]{Kaluzny+1987}Kaluzny, J., \& Shara, M.~M., 1987, {\it ApJ}, {\bf314}, 585, \OEJVbibcode{1987ApJ...314..585K}
\bibitem[Kenyon \& Hartmann(1995)]{Kenyon+1995}Kenyon, S.~J., \& Hartmann, L., 1995, {\it ApJS}, {\bf101}, 117\OEJVbibcode{1995ApJS..101..117K}
\bibitem[Kwee \& van Woerden(1965)]{Kwee+1965} Kwee, K.K., \& van Woerden, H., 1956, {\it BAN}, {\bf12}, 327\OEJVbibcode{1956BAN....12..327K}
\bibitem[Lipunov et~al.(2010)]{Lipunov+2010}Lipunov, V., Kornilov, V., Gorbovskoy, E., et~al., 2010, {\it Advances in Astronomy}, {\bf2010}, \OEJVbibcode{2010AdAst2010E..30L}
\bibitem[McClure(1974)]{McClure1974}McClure, R.~D., 1974, {\it ApJ}, {\bf194}, 355, \OEJVbibcode{1974ApJ...194..355M}
\bibitem[Mochejska et~al.(2008)]{Mochejska+2008}Mochejska, B.~J., Stanek, K.~Z., Sasselov, D.~D., et~al., 2008, {\it ACTAA}, {\bf58}, 263, \OEJVbibcode{2008AcA....58..263M}
\bibitem[Pickles(1998)]{Pickles1998}Pickles, A.~J., 1998, {\it PASP}, {\bf110}, 863, \OEJVbibcode{1998PASP..110..863P}
\bibitem[Platais et~al.(2003)]{Platais+2003}Platais, I., Kozhurina-Platais, V., Mathieu, R.~D., Girard, T.~M., \& van Altena, W.~F., 2003, {\it AJ}, {\bf126}, 2922, \OEJVbibcode{2003AJ....126.2922P}
\bibitem[Roeser et~al.(2008)]{Roeser+2008}Roeser, S., Schilbach, E., Schwan, H., et~al., 2008, {\it VizieR Online Data Catalog}, {\bf1312}, 0, \OEJVbibcode{2008yCat.1312....0R}
\bibitem[Rose \& Hintz(2007)]{Rose+2007}Rose, M.~B., \& Hintz, E.~G., 2007, {\it AJ}, {\bf134}, 2067, \OEJVbibcode{2007AJ....134.2067R}
\bibitem[Sarajedini et~al.(1999)]{Sarajedini+1999}Sarajedini, A., von Hippel, T., Kozhurina-Platais, V., \& Demarque, P., 1999, {\it AJ}, {\bf118}, 2894, \OEJVbibcode{1999AJ....118.2894S}
\bibitem[Skrutskie et~al.(2006)]{Skrutskie+2006}Skrutskie, M.~F., Cutri, R.~M., Stiening, R., et~al., 2006, {\it AJ}, {\bf131}, 1163, \OEJVbibcode{2006AJ....131.1163S}
\bibitem[Stetson et~al.(2004)]{Stetson+2004}Stetson, P.~B., McClure, R.~D., \& VandenBerg, D.~A., 2004, {\it PASP}, {\bf116}, 1012, \OEJVbibcode{2004PASP..116.1012S}
\bibitem[Tody(1986)]{Tody1986}Tody, D., 1986, {\it SPIE proccedings}, {\bf627}, 733, \OEJVbibcode{1986SPIE..627..733T}
\bibitem[Twarog et~al.(1997)]{Twarog+1997}Twarog, B.~A., Ashman, K.~M., \& Anthony-Twarog, B.~J., 1997, {\it AJ}, {\bf114}, 2556, \OEJVbibcode{1997AJ....114.2556T}
\bibitem[Zhang et~al.(2002)]{Zhang+2002}Zhang, X.~B., Deng, L., Tian, B., \& Zhou, X., 2002, {\it AJ}, {\bf123}, 1548, \OEJVbibcode{2002AJ....123.1548Z}
\bibitem[Zhang et~al.(2004)]{Zhang+2004}Zhang, X.~B., Deng, L., Zhou, X., \& Xin, Y., 2004, {\it MNRAS}, {\bf355}, 1369, \OEJVbibcode{2004MNRAS.355.1369Z}

\end{thebibliography}
\end{document}